\documentclass[10pt, aps, pra, nofootinbib, 
               amsmath, amssymb, twocolumn,
               preprintnumbers, showpacs,
               raggedbottom,
               floatfix]{revtex4-1}

\usepackage[T1]{fontenc}
\usepackage{textcomp}

\usepackage{etoolbox}
\usepackage{etextools}
\usepackage[acronym]{glossaries}
\glsdisablehyper
\makeglossaries

\usepackage[utf8]{inputenc}

\usepackage[breaklinks]{hyperref}
\usepackage{graphicx}
\usepackage{dcolumn}
\usepackage{calc}

\usepackage[tracking]{microtype}
\SetTracking{ encoding = *, shape = sc }{ 25 }

\AtBeginDocument{\usepackage{bm}}

\usepackage{siunitx}
\usepackage[version=3]{mhchem}

\makeatletter
\newcommand\na[3][\@empty]{%
  {%
    \ifx#1\@empty
      \lowercase{\def\short{\protect\mysc{#2}}}
    \else
      \def\short{#1}
    \fi
    \expandnext{\newacronym{#2}}{\short}{#3}
  }
}
\let\mysc\textsc
\makeatother
\na{TD}{time-dependent}
\na{BEC}{Bose-Einstein condensate}
\na{BCS}{Bardeen-Cooper-Schrieffer}
\na{FFT}{fast Fourier transform}
\na[\textsc{b}d\textsc{g}]{BdG}{Bogoliubov-de Gennes}
\na{DFT}{density functional theory}
\na{SLDA}{superfluid local density approximation}
\na{TDSLDA}{time-dependent superfluid local density approximation}
\na{TDDFT}{time-dependent density functional theory}
\na{ASLDA}{asymmetric \gls{SLDA}}
\na{QMC}{quantum Monte Carlo}
\na[\textsc{d}\textup{o}\textsc{e}]{DoE}{Department of Energy}
\na{LANL}{Los Alamos National Laboratory}
\na{LDRD}{}
\na{NCCS}{}

\renewcommand{\d}{\mathrm{d}}
\providecommand{\abs}[1]{\lvert{#1}\rvert}

\providecommand{\vect}[1]{\vec{#1}}

\newcommand{\I}{\mathrm{i}}
\DeclareMathOperator{\Tr}{Tr}

\providecommand{\spinsum}{\!\!\!\sum_{\sigma\in\{\uparrow,\downarrow\}}\!\!\!\!}
\providecommand\intd[2]{\int\d^{#1}{#2}\;}
\providecommand\bra[1]{\langle #1|}
\providecommand\ket[1]{|#1\rangle}
\providecommand\braket[1]{\langle #1\rangle}
\providecommand\op[1]{\hat{#1}}
\providecommand\mat[1]{\bm{#1}}

\providecommand\pdiff[2]{\frac{\partial #1}{\partial #2}}

\usepackage{lettrine}
\usepackage[sc,osf]{mathpazo}
\usepackage[euler-digits,small]{eulervm}
\AtBeginDocument{\renewcommand{\hbar}{\hslash}}

\LettrineOptionsFor{T}{
  lines=3,
  loversize=0.1,
  lraise=-0.03,
  lhang=0.49,
  findent=0.4em,
  nindent=-0.0\LettrineWidth-0.5em}

\LettrineOptionsFor{U}{
  lines=3,
  loversize=0.05,
  lraise=0.03,
  lhang=0.11,
  findent=-0.1em,
  nindent=0.2em}

\LettrineOptionsFor{U}{
  lines=3,
  loversize=0.09,
  lraise=0.01,
  lhang=0.13,
  findent=-0.4em,
  nindent=0.4em,
  slope=0.5em}

\LettrineOptionsFor{L}{
  lines=3,
  loversize=0.07,
  lraise=0.0,
  lhang=0.19,
  findent=-1.0em,
  nindent=0.6em,
  slope=0.8em}

\renewcommand{\figurename}{Figure}
\renewcommand{\tablename}{Table}
\makeatletter
\renewcommand{\fnum@figure}{\textbf{\figurename~\thefigure}}
\renewcommand{\fnum@table}{\textbf{\tablename~\thetable}}
\makeatother

\usepackage[utf8]{inputenc}

\begin{document}

\title{Time-Dependent Superfluid Local Density Approximation}

\author{%
  Aurel Bulgac,\textsuperscript{1}
  Michael McNeil Forbes,\textsuperscript{1,2}}

\affiliation{\textsuperscript{1}Department of Physics,
  University of Washington, Seattle, WA 98195--1560, USA,}
\affiliation{\textsuperscript{2}Institute for Nuclear Theory, 
  University of Washington, Seattle, Washington 98195--1550 USA}

\providecommand{\MMFGRANT}{\textsc{de-fg02-00er41132}}
\providecommand{\ABGRANTa}{\textsc{de-fg02-97er41014}}
\providecommand{\ABGRANTb}{\textsc{de-fc02-07er41457}}

\begin{abstract}
  \noindent
  The \gls{TDSLDA} is an extension of the Hohenberg-Kohn \gls{DFT} to
  time-dependent phenomena in superfluid fermionic systems.  Unlike linear
  response theory, which is only valid for weak external fields, the
  \gls{TDSLDA} approach allows one to study non-linear excitations in fermionic
  superfluids, including large amplitude collective modes, and the response to
  strong external probes.  Even in the case of weak external fields, the
  \gls{TDSLDA} approach is technically easier to implement.  We will illustrate
  the implementation of the \gls{TDSLDA} for the unitary Fermi gas, where
  dimensional arguments and Galilean invariance simplify the form of the
  functional, and \textit{ab initio} input from \gls{QMC} simulations
  fix the coefficients to quite high precision.
\end{abstract}
\preprint{\textsc{int-pub-12-056}}

\maketitle
\glsresetall

%\section{Introduction}\noindent
\lettrine{L}{inear response theory} is a popular tool for studying the dynamics
of a quantum many-body system.  Formally, the change in the number density
(often referred to as the transition density) in response to a weak external
potential $V_{\text{ext}}(\vect{r},t)$ is given by
\begin{equation}
  \label{bulgac_eq:linres_t}
  \delta n(\vect{r},t)
  = \int\d\vect{r}'\d{t}'\Pi(\vect{r},t,\vect{r}',t')
  V_{\text{ext}}(\vect{r}',t')
\end{equation}
where $\Pi(\vect{r},t,\vect{r}',t')$ is the linear response function of the
system. Since, for a system in equilibrium, $\Pi(\vect{r},t,\vect{r}',t')$
depends only on the difference $t-t'$, one usually works with the Fourier
transforms:
\begin{equation}
  \label{bulgac_eq:linres_o}
  \delta n(\vect{r},\omega)
  = \int\d\vect{r}'\Pi(\vect{r},\vect{r}',\omega)
  V_{\text{ext}}(\vect{r}',\omega). 
\end{equation}
The linear response function $\Pi(\vect{r},\vect{r}',\omega)$ has poles at
frequencies corresponding to the various excited states of the system, which
allows one to express these excited states in a form independent of the external
probe:
\begin{equation}
  \label{bulgac_eq:linres_L}
  \int \d\vect{r}'\Lambda(\vect{r},\vect{r}',\omega) \delta n(\vect{r}',\omega) 
  =0.
\end{equation}
Here $\Lambda(\vect{r},\vect{r}',\omega)$ is the operator inverse of
$\Pi(\vect{r},\vect{r}',\omega)$. The existence of
$\Lambda(\vect{r},\vect{r}',\omega)$ is nontrivial as the operator
$\Pi(\vect{r},\vect{r}',\omega)$ may be singular due to zero modes
(Goldstone modes) arising from various conservation laws.

This approach is appealing, because solutions of Eq.~(\ref{bulgac_eq:linres_L})
describe intrinsic excitations of the system.  However, it is clearly limited to
describing small amplitude excitations where the response remains linear and the
external potential is weak.  These equations are also technically difficult to
solve due to the high dimensionality of the matrices involved: especially in the
case of inhomogeneous systems.  This makes it practically impossible to
implement a fully three-dimensional calculation, and they have only been solved
in systems with a high degree of symmetry: infinite homogeneous systems for
example, or axially/spherically symmetric configurations. Even in such cases,
limiting assumptions or approximations are often required.

Here we shall describe a different approach: \gls{TDDFT}.  This not only allows
one to study non-linear excitations, but also allows one to consider fully
three-dimensional equations.  Although exact in principle, there is no simple
prescription for computing an exact density functional in a non-perturbative
theory (see~\cite{drut_furnstahl_10} for recent discussions), and one must first
formulate an approximate functional that captures the relevant physics.  In the
case of the unitary Fermi gas, the lack of scales greatly restricts the possible
forms for the functional, and an extremely simple form --- the
\gls{SLDA}~\cite{bulgac_07} (described in Sec.~\ref{bulgac_sec:functional}) ---
appears to capture much of the relevant physics.  The \gls{TDSLDA} requires one
to solve a system of coupled time-dependent three-dimensional nonlinear
Schr\"{o}dinger-like equations of the form
\begin{equation}
  \label{bulgac_eq:gen}
  \I\hbar\pdiff{\vect{\Psi}_k(\vect{r},t)}{t} = 
   [\op{H}(\vect{r},t) + \op{V}_{\text{ext}}(\vect{r},t)]
   \vect{\Psi}_k(\vect{r},t).
\end{equation}
Here $\vect{\Psi}_k(\vect{r},t)$ is a vector of single-quasiparticle
wavefunctions, the exact meaning of which will be explained below, and the
corresponding single-particle Hamiltonian $\op{H}(\vect{r},t)$ is a partial
differential operator. The main complexity of this system of equations arises
from the fact that the single-particle Hamiltonian $\op{H}(\vect{r},t)$ depends
non-linearly on all the single-quasiparticle wavefunctions
$\vect{\Psi}_k(\vect{r},t)$.  The simplification is that $\op{H}$ contains only
differential operators (no integral operators either in time or space), and can
be efficiently applied on each wavefunction independently, allowing the method
to be efficiently parallelized.  Since no matrix operations are involved (the
kinetic and potential parts are applied separately and efficiently using the
\gls{FFT}, and the memory requirements are significantly reduced compared to
solving Eq.~(\ref{bulgac_eq:linres_L}).

The \gls{TDSLDA} also has conceptual advantages over some traditional approaches
to superfluid dynamics: unlike two-fluid hydrodynamics, the \gls{TDSLDA} can
correctly describe \emph{quantized} vortices and their dynamics, and contains
naturally the critical flow velocity at which a superfluid can turn into a
normal fluid; in contradistinction to the Gross-Pitaevskii or Ginzburg-Landau
approaches, the normal fluid to superfluid transition is within the scope of the
theory.  Moreover, a number of large amplitude collective modes have been
studied with the \gls{TDSLDA} that defy a description within two-fluid
hydrodynamics, Ginzburg-Landau, or Gross-Pitaevskii
frameworks~\cite{bulgac_yoon_09}.

\section{Methodology}\noindent
A precise formal statement of a \gls{DFT} starts with some physically motivated
energy functional $E[n_1, n_2, \cdots]$ of various densities $n_i(\vect{r}, t)$.
To simplify the formal structure, we express this as a function of the density
matrix $E(\op{\rho})$ though in the end we shall only consider local functions
(see Sec.~\ref{bulgac_sec:functional}).  Once specified, one simply minimizes
the free energy $F(\op{\rho}) = E(\op{\rho}) + T\Tr(\op{\rho}\ln\op{\rho})$
subject to the normalization constraint on $\op{\rho} +
\mat{C}\op{\rho}^{T}\mat{C} = \mat{1}$ dictated by Fermi statistics, where
$\mat{C} = \mat{C}^{T}$ is the charge conjugation matrix.  The constrained
minimization of the functional $F(\op{\rho})$ results in the standard Fermi
distribution\footnote{Formally, this constraint can be implemented using a
  Lagrange multiplier, but it is much easier to see the results by letting
  $\op{\rho} = \mat{1}/2 + \mat{x} - \mat{C}\mat{x}^T\mat{C}$ where $\mat{x}$ is
  unconstrained, and then performing the variation with respect to $\mat{x}$.}
\begin{equation}
  \label{bulgac_eq:Formal_Fermion}
  \op{\rho} = \sum_{k}\ket{k}n_{FD}(E_{k})\bra{k}
  =
  \frac{1}
  {1+e^{\beta \left(\mat{H}(\op{\rho}) - \mat{C}\mat{H}^{T}(\op{\rho})\mat{C}\right)}},
\end{equation}
to obtain the following equations of motion
\begin{subequations}
  \label{bulgac_eq:KS_1}
  \begin{align}
    \op{H}(\op{\rho})\ket{k} 
    &= \frac{\delta E(\op{\rho})}{\delta\op{\rho}}\ket{k} =
    E_{k}\ket{k}, \\
    \op{\rho} &=
    \sum_{k}\ket{k}n_{FD}(E_k)\bra{k},
  \end{align}
\end{subequations}
which must be solved self-consistently. The eigenvalues
$E_{k}$ are the Lagrange multipliers of the associated
normalization constraint.  The formulation of the \gls{TDDFT} follows simply by
using $\op{H}(\op{\rho})$ to generate the time evolution of the single particle
states,
\begin{equation}
  \label{bulgac_eq:KD_TD}
  \I\hbar\partial_{t} \ket{k} = \op{H}_{t}(\op{\rho})\ket{k},
\end{equation}
typically in the presence of some time-dependent external potential included in
$\op{H}_{t}(\op{\rho}) = \op{H}(\op{\rho}) + V_{\text{ext}}(t)$, for
example, or a gauge coupling in the case of an electromagnetic external field.

The physical content of the \gls{DFT} enters through the formulation of the
function $E(\op{\rho})$ as we shall discuss in Sec.~\ref{bulgac_sec:functional}.
The technical challenges are: 1) diagonalizing the single-particle
Hamilton~(\ref{bulgac_eq:KS_1}); 2) solving the self-consistency equations to
determine stationary (ground state) configurations; and 3) stably and
self-consistently evolving the single-particle states~(\ref{bulgac_eq:KD_TD}) to
describe the dynamics.  Typically one applies all three techniques, first
solving for an initial stationary configuration, then driving the system to
explore the dynamics --- stirring to generate vortices for example.

\subsection{The Functional}\noindent
\label{bulgac_sec:functional}
In practice, one does not work explicitly with the density matrix $\op{\rho}$
but rather with a set of physically motivated local densities.  It is convenient
to express these concepts in the language of second quantization.  We consider
two species with operators $\op{c}_{\uparrow}$ and $\op{c}_{\downarrow}$
representing two hyperfine states.

For a two component system, the most general wavefunction that allows for all
possible pairings has four components: $\smash{\vect{\op{\Psi}} =
  (\op{c}_{\uparrow}, \op{c}_{\downarrow}, \op{c}_{\uparrow}^{\dagger},
  \op{c}_{\downarrow}^{\dagger})}$.  In terms of components of the wavefunction,
we will write $\smash{\mat{H} \vect{\Psi}_{k} = E_{k}\vect{\Psi}_{k}}$ where:
$\vect{\Psi}_{k}(\vect{r}, t) = \braket{\vect{r}|k} =
\bigl(u_{k\uparrow}(\vect{r}, t), u_{k\downarrow}(\vect{r}, t),
v_{k\uparrow}(\vect{r}, t), v_{k\downarrow}(\vect{r}, t)\bigr)$.  In what
follows we shall drop the explicit $(\vect{r}, t)$ dependence.  In this
formulation, the time evolution of a single-particle wavefunction
$\vect{\Psi}_{k}$ is:
\begin{multline}
  \label{bulgac_eq:tddft}
  \I \hbar \pdiff{}{t}
  \begin{pmatrix}
    u_{k\uparrow} \\ 
    u_{k\downarrow} \\ 
    v_{k\uparrow} \\ 
    v_{k\downarrow}
  \end{pmatrix} \\
  =
  \begin{pmatrix}
    {h_{\uparrow}+U_{\uparrow}}\!\!\! & \chi & 0 & \Delta \\
    \chi^{*} & \!\!\!{h_{\downarrow}+U_{\downarrow}}\!\!\!  & -\Delta & 0 \\
    0 & -\Delta^*&  \!\!\!{-h_{\uparrow}^*-U_{\uparrow}}\!\!\! & -\chi^* \\
    \Delta^*& 0 & -\chi & \!\!\!{-h_{\downarrow}^*-U_{\downarrow}}
  \end{pmatrix}
  \begin{pmatrix}
    u_{k\uparrow} \\ 
    u_{k\downarrow} \\ 
    v_{k\uparrow} \\ 
    v_{k\downarrow}
  \end{pmatrix}
\end{multline}
where $h_{\uparrow,\downarrow} = -\nabla^2/(2m_{\uparrow,\downarrow})$,
$U_{\uparrow,\downarrow}$ is the self-energy, and $\Delta \propto
\braket{\op{c}_{\uparrow}^\dagger\op{c}^{\dagger}_{\downarrow}}$ is the pairing
field.  One needs this full four-component formalism if $\chi \propto
\smash{\braket{\op{c}_{\uparrow}^\dagger\op{c}_{\downarrow}}\neq 0}$.  (A spin-orbit
coupling in the nuclear problem would appear here for example.)  For the unitary
gas, however, we consider only attractive $s$-wave interactions (thus, $\chi =
0$), allowing us to express everything in terms of the usual two-component
\gls{BdG} form $\vect{\Psi}_k = (u_k, v_k)$:
\begin{equation}
  \label{bulgac_eq:tddft_2d}
  \I \hbar \pdiff{}{t}
  \begin{pmatrix}
    u_k \\ 
    v_k
  \end{pmatrix} = 
  \begin{pmatrix}
    h_{\uparrow}+U_{\uparrow} & \Delta \\
    \Delta^*& -h_{\downarrow}^*-U_{\downarrow}
  \end{pmatrix}
  \begin{pmatrix}
    u_k \\ 
    v_k
  \end{pmatrix}.
\end{equation}
Note that the structure of these equations is that of a single quasiparticle
Hamiltonian: Indeed, for the choice of functional we consider below, this will
look formally like the standard \gls{BdG} equations, however, the coefficients
will be determined from the functional rather than from a direct mean-field
approximation of a microscopic theory.  In the usual formulation of a \gls{DFT}
for normal systems, the single particle states need not bear any formal
relationship to the physical quasiparticles.  Within the \gls{SLDA}, however, we
have found that the quasiparticle properties --- their dispersion relationship
for example --- can also be successfully modeled with the appropriate choice of
functional.

For simplicity we shall consider here only the symmetric case $n_{\uparrow} =
n_{\downarrow} = n_{+}/2$ where the two states have identical masses and
describe the \gls{SLDA}.  (See~\cite{bulgac_forbes_chapter_11} for details about
the \gls{ASLDA} extension.)  We consider three densities and one current:
\begin{align}
  n_{+}(\vect{r}) &= { 2\sum_{k}\abs{v_{k}(\vect{r})}^2 n_{FD}(-E_k)}
  \sim \spinsum\braket{\op{c}_{\sigma}^\dagger\op{c}_{\sigma}}
  ,\label{bulgac_eq:Densities}\\
  \tau_{+}(\vect{r}) &= { 2\sum_{k}\abs{\vect{\nabla} v_{k}(\vect{r})}^2
    n_{FD}(-E_k) } 
  \sim \!\!\spinsum\braket{
    \vect{\nabla}\op{c}_{\sigma}^\dagger\cdot\vect{\nabla}\op{c}_{\sigma}
  },\nonumber\\
  \nu(\vect{r}) 
  &= \tfrac{1}{2}\!\sum_{k} u_{k}(\vect{r})v_{k}^{*}(\vect{r})
  \Bigl[
  n_{FD}(-E_k) 
  - n_{FD}(E_k)\Bigr] 
  \!\sim \braket{\op{c}_{\uparrow}\op{c}_{\downarrow}},\nonumber\\
  \vect{j}_{+}(\vect{r}) &=
  i\sum_{k} \left[v^*_{k}(\vect{r}) \vect{\nabla} v_{k}(\vect{r})  
    - v_{k}(\vect{r}) \vect{\nabla} v^*_{k}(\vect{r})\right]
  n_{FD}(-E_k).\nonumber
\end{align}
We use the kinetic energy density $\tau_{+}$ in the spirit of Kohn-Sham, and the
anomalous density $\nu$ to account for pairing within a local theory.  For
time-reversal invariant ground states, the current density $\vect{j}_{+}$
vanishes.  It must be considered when considering time-dependence to ensure that
the energy density is covariant under local Galilean transformations.  In
nuclear physics Galilean invariance have been considered for quite some
time~\cite{engel_brink_75, dobaczewski_dudek_95, nesterenko_kleinig_08,
  bender_heenen_03}, and the contribution of these currents is often essential
for describing the properties of excited states.  It is easily demonstrated
(see~\cite{bulgac_forbes_chapter_11} for details) that when changing to a frame
with velocity $\vect{v}$, the currents and kinetic densities transform as
\begin{align}
  \label{bulgac_eq:Galilean_j}
  \vect{j}_{+} &\rightarrow
  \vect{j}_{+} + M\vect{v}n_{+}, &
  \tau_{+} &\rightarrow
  \tau_{+} + \vect{v}\cdot\vect{j}_{+} 
  + \tfrac{1}{2}M\abs{\vect{v}}^2n_{+}
\end{align}
where $M = M_{\uparrow} = M_{\downarrow}$ is the bare mass of the particles.  It
follows that for symmetric two-component systems, the following is
Galilean invariant:
\begin{equation}
  \label{bulgac_eq:kincur}
  \tilde{\tau}_{+} = \tau_{+} 
  - \frac{\abs{\vect{j}_{+}}^2}{2M n_{+}}.
\end{equation}
The center of mass motion may be separated from the intrinsic energy density
(the total energy $E = \intd[3]{\vect{r}}\mathcal{E}$)
\begin{equation}
  \mathcal{E} = \frac{\abs{\vect{j}_{+}}^2}{2n_{+}}
  + \tilde{\mathcal{E}}(\tilde{\tau}_{+}, n_{+}, \nu)
\end{equation}
such that $\tilde{\mathcal{E}}$ is locally Galilean invariant.

The form of the functional is further restricted by the fact that the anomalous
density $\nu(\vect{r},\vect{r}') \sim
\braket{\op{c}_{\uparrow}(\vect{r})\op{c}_{\downarrow}(\vect{r}')} \propto
\abs{\vect{r} - \vect{r}'}^{-1}$ is ultraviolet divergent in the local
approximation.  This divergence also appears in the kinetic term $\tau_{+}$ and
the two always enter the functional as
\begin{equation}
  \alpha\frac{\tilde{\tau}_{+}}{2} 
  + \frac{\nu^\dagger\nu}{n_{+}^{1/3}/\gamma - \Lambda/\alpha},
\end{equation}
where $\gamma$ parametrizes the pairing strength, $\alpha=M/M_{\text{eff}}$ is
the inverse effective mass, and $\Lambda$ is a momentum space cutoff.  The most
straight-forward functional constructed from these quantities is the \gls{SLDA}:
\begin{multline}
  \tilde{\mathcal{E}}_{\gls*{SLDA}}(\tau_{+}, n_{+}, \nu) =
  \frac{\hbar^2}{M}
  \Biggl(
    \left[
      \alpha\frac{\tilde{\tau}_{+}}{2} +
      \frac{\nu^\dagger\nu}{n_{+}^{1/3}/\gamma - \Lambda/\alpha}
    \right] +\\
    + \beta \frac{3}{10}(3\pi^2)^{2/3}n_{+}^{5/3}
  \Biggr).
\end{multline}
Varying this functional leads to the following identification of the
single particle Hamiltonian $h = h_{\uparrow} = h_{\downarrow}$, potential $U$,
and gap parameter $\Delta$:
\begin{align}
  \label{bulgac_eq:potentials}
  h &= -\alpha\frac{\hbar^2\nabla^2}{2M} - \mu,\\
  \Delta &= -\frac{\nu}{n_{+}^{1/3}/\gamma - \Lambda/\alpha},\\
  U &= \beta \frac{\hbar^2}{2M}(3\pi^2)^{2/3}n_{+}^{2/3} - 
  \frac{\Delta^\dagger\Delta}{3\gamma n_{+}^{2/3}} + V_{\text{ext}}.
\end{align}
For spatially varying systems, momentum is not a good quantum number and a
simple momentum space cutoff cannot be implemented.  Instead, one can use an
energy cutoff, limiting the sums in Eqs.~(\ref{bulgac_eq:Densities}) for
energies $\abs{E_{k}} < E_c$.  The homogeneous equations can then be used to
translate this into a position dependent $\Lambda(\vect{r})$ that may be used in
the previous equations and which has very good convergence
properties~\cite{bulgac_yu_02} (see also~\cite{bulgac_forbes_chapter_11}):
\label{bulgac_eq:reg}
\begin{gather}
  \Lambda_c(\vect{r}) = 
  \frac{M}{\hbar^2}
  \frac{k_c(\vect{r})}{2\pi^2}
  \left\{
    1 - \frac{k_0(\vect{r})}{2k_c(\vect{r})}
    \ln\frac{k_c(\vect{r}) + k_0(\vect{r})}{k_c(\vect{r}) - k_0(\vect{r})}
  \right\},\\
  \intertext{where $k_0$ and $k_c$ are defined by}
  \begin{aligned}
    \alpha\frac{\hbar^2k_0^2(\vect{r})}{2M}  -
    \mu + U(\vect{r})
    &= 0, &
    \alpha\frac{\hbar^2 k_c^2(\vect{r})}{2M} -
    \mu + U(\vect{r}) 
    &= E_c.
  \end{aligned}\nonumber
\end{gather}
To complete the functional, one must determine the parameters $\alpha$, $\beta$,
and $\gamma$.  We do this by matching the predictions of the functional in the
thermodynamic limit to accurate \gls{QMC} calculations.  Fitting the
energy and quasiparticle spectrum determines the following values for the
unitary gas (see~\cite{bulgac_forbes_chapter_11} for a detailed discussion of
this fitting procedure):
\begin{align*}
  \alpha &= \num{1.094(17)},&
  \beta &= \num{-0.526(18)},&
  \gamma^{-1} &= \num{-0.0907(77)}.
\end{align*}
The \gls{TDSLDA} satisfies all expected conservation laws: energy in the absence
of time-dependent fields, linear/angular momentum if the corresponding
symmetries are not broken, gauge and Galilean invariance, and particle number in
the absence of applied external pairing fields.

\subsection{Technical Notes}\noindent
\label{bulgac_sec:technical_methods}
Solving the self-consistency conditions requires solving such a large number of
simultaneous equations that typical root finding methods employing a Jacobian
computation are prohibitive.  However, treated as an iterative method --- take
an initial set of densities, form the potential~(\ref{bulgac_eq:potentials}),
diagonalizing the Hamiltonian to obtain a new set of single particle wave
functions, and then form a new set of densities~(\ref{bulgac_eq:Densities}) ---
the self-consistency cycle is typically close to convergent.  As a result, a
memory limited implementation of Broyden's method~\cite{baran_bulgac_08} works
well to accelerate convergence, thereby determining equilibrium configurations
to use as an initial state for a subsequent time-dependent simulation.

The output of this is a complete set of wavefunctions, typically represented on
a periodic lattice.  These are then fed into the time dependence
equations~(\ref{bulgac_eq:KD_TD}) to generate the time-dependent states
$\ket{n(t)}$.  Note that at each time-step, the Hamiltonian must be updated to
reflect the current ensemble of states.  We have found that a multistep
predictor-modifier-corrector method due to Adams-Bashforth-Milne (see
\cite{hamming_book_73}) works well (see~\cite{bulgac_roche_08} for
implementation details and parallel scaling performance.)  Periodic lattices
enable us to use the \gls{FFT} to efficiently transform the wave functions
between position and momentum space so that the kinetic and potential parts of
the Hamiltonian may be applied by simple multiplication. This allows us to
perform fully three-dimensional simulations.

\begin{figure*}[t]
  \includegraphics*[width=0.5\textwidth]{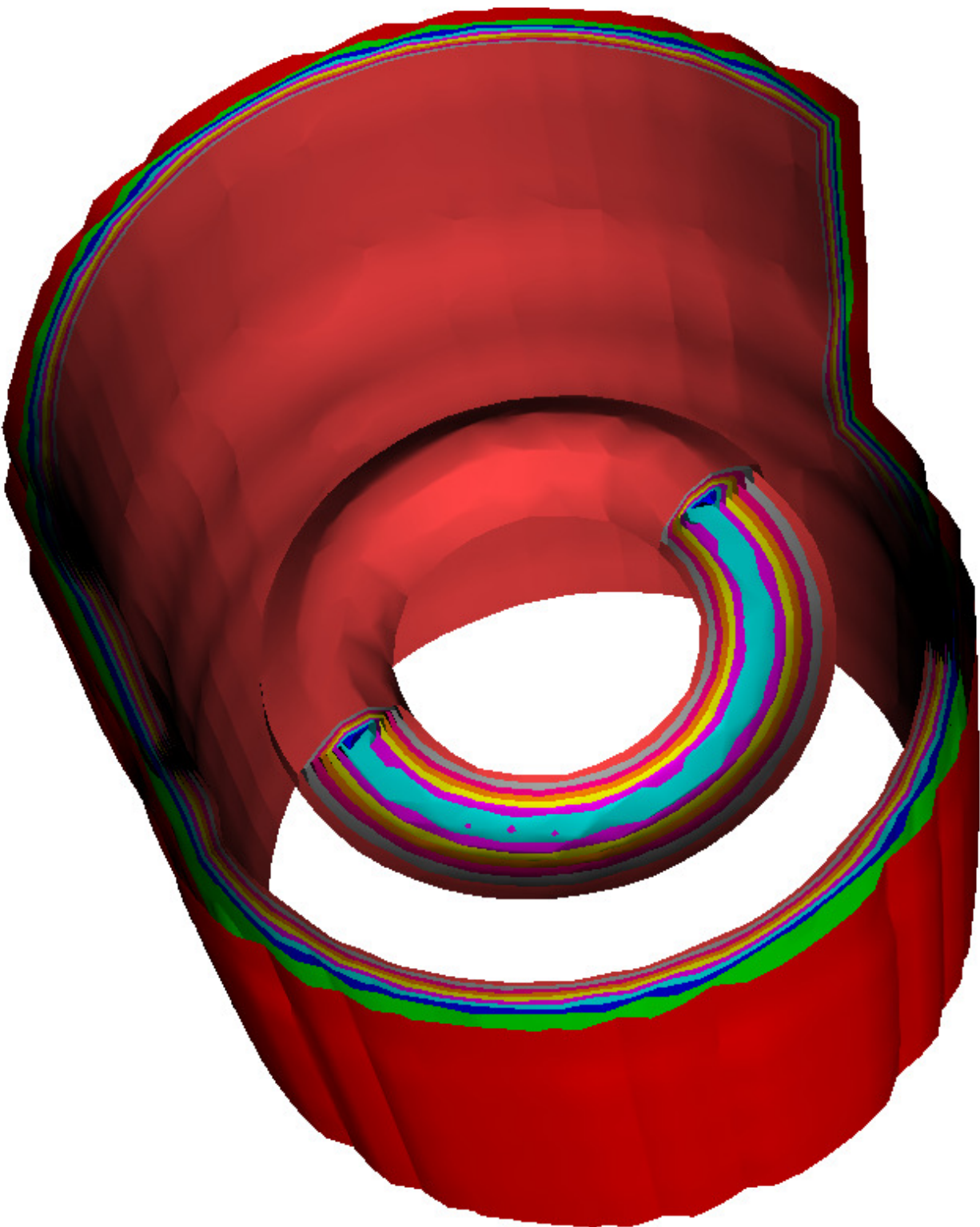}%
  \includegraphics*[width=0.5\textwidth]{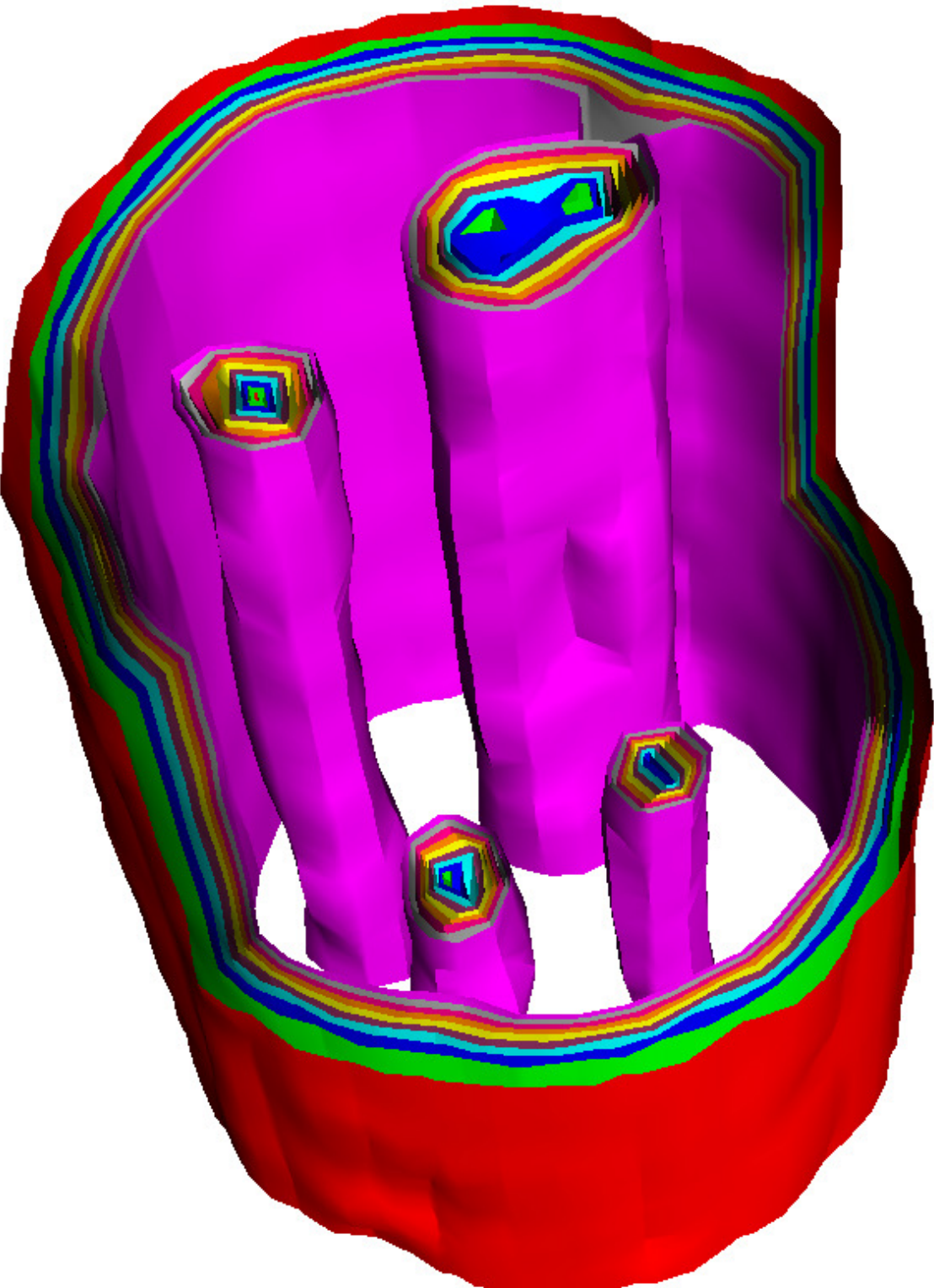}
  \caption{Two frames of 3\textsc{d} time dependent simulations of a unitary
    Fermi gas confined to a cylindrical trap and subject to a time dependent
    external potential.  On the left, a hard sphere moved along the trap axis,
    generating a vortex ring in its wake.  On the right, the external potential
    was a vertical rod and a diametrically opposed sphere which stirred the
    system, generating five vortices.  Kelvin waves have been excited along each
    vortex. The last two vortices have been generated simultaneously: they are
    essentially on top of each other and separate at a later time.
  }\label{bulgac_fig:3d}
\end{figure*}

\subsection{Validity domain}\noindent
If one has an exact density functional, then the \gls{TDDFT} technique can be
shown to deliver the exact time-evolution of the one-body density~\cite{
  rajagopal_callaway_73, peuckert_78, runge_gross_84}.  If one is interested in
higher-order operators, however, then extensions to the technique are
required~\cite{bulgac_10}.  These are significantly more costly, but still
within computational reach for carefully chosen problems.

The main limitation is that an exact density functional is not known.  Thus, the
\gls{DFT} requires careful benchmarking to determine the domain of validity.  At
present, the \gls{SLDA} has been formulated and fit using \gls{QMC} calculations
of the $T=0$ thermodynamic limit of the three-dimensional unitary Fermi gas.
This has been benchmarked against trapped systems to an accuracy of a few
percent~\cite{bulgac_forbes_chapter_11}, indicating that the omitted gradient
corrections are quite small.  Thus, the \gls{SLDA} is reliable for cold
symmetric systems up to small gradients corrections.  The asymmetric extension
(the \gls{ASLDA}) has also been formulated and fit to \gls{QMC} data.  The
extension to finite-temperatures is still an open problem.

\subsection{Relevance to other theories}\noindent
The \gls{ASLDA} subsumes the usual mean-field \gls{BdG} equations, but extends
these considerably. For example, it includes a self-energy contribution that is
neglected in the zero-range limit of the mean-field \gls{BdG} equations.  The
\gls{ASLDA} lacks the variational property of the mean-field \gls{BdG}
equations, but with careful validation, has the ability to provide a much more
quantitatively accurate description of fermionic
superfluids~\cite{bulgac_forbes_chapter_11}.

\section{Applications}\noindent
We present here briefly two quite spectacular results obtained using the
\gls{TDSLDA} in a unitary gas. We prepare a system in its ground state in an
axially symmetric trap (with an essentially flat bottom) and homogeneous with
periodic boundary conditions in the third direction.  We then adiabatically
introduce two types of quantum stirrers: 1) a spherical projectile flying along
the symmetry axis with a speed $v_p=0.2\,v_F$ (where $v_F$ is the Fermi
velocity); and 2) a rod parallel to the symmetry axis with a diametrically
opposed sphere (breaking translational invariance along the tube) moving with a
constant angular velocity about the center of the tube and a linear velocity
lower than the critical velocity of the unitary Fermi gas $v_c\approx
0.365\,v_F$~\cite{sensarma_randeria_06, combescot_kagan_06}. In the first case,
the spherical projectile, after passing through the system, generates a rather
elusive excitation mode of a superfluid: a vortex ring.  In the second case, the
two quantum stirrers (the rod and the sphere) generate five vortices.  The
sphere breaks the translational symmetry, exciting Kelvin modes along the
vortices, and, at the same time, exciting phonons in the superfluid to form a
complicated mixture of dynamical modes.  In each of these simulations we solved
about \num{22000} time-dependent 3\textsc{d} coupled nonlinear partial
differential equations on a $32^3$ spatial lattice for a sufficiently long
period of time.

\section{Relevance to Other Systems}\noindent
Even though we have only illustrated the \gls{TDSLDA} in the case of a unitary
Fermi gas, this is a rather general approach suitable to describe the dynamics
of virtually any fermionic superfluid with $s$-wave pairing. The \gls{TDSLDA}
has already been used to describe nuclear systems: in particular, the first
attempt to describe induced nuclear fission was recently performed.  Although
not yet explored, it appears that the extension to pairing in other partial
waves ($p$-wave and $d$-wave for example) is straightforward.

\glsunset{LDRD}
\glsunset{NCCS}
\section*{Acknowledgments}\noindent
We acknowledge numerous discussions with our collaborators Y.-L. Luo,
P. Magierski, K.J. Roche, S. Yoon, Y. Yu, and funding from the \gls{DoE} under
grants \ABGRANTa, \ABGRANTb, \MMFGRANT, and the \gls{LDRD} program at
\gls{LANL}. Calculations reported here have been performed on the
Jaguar\textsc{pf} supercomputer (Cray \textsc{xt5}, \gls{NCCS}).

%\bibliographystyle{apsrev4-1}
%\bibliography{bulgac}

\end{document}